\documentclass[%
reprint,
superscriptaddress,
 amsmath,amssymb,
 aps,
 pra,
]{revtex4-2}

\usepackage[utf8]{inputenc}
\usepackage[T1]{fontenc}
\usepackage[english]{babel}

\usepackage{dsfont}
\usepackage[colorlinks=true, citecolor=blue,urlcolor=blue,linkcolor=blue,filecolor=black]{hyperref}
\usepackage[colorinlistoftodos, color=green!40, prependcaption]{todonotes}

\usepackage{graphicx}

\usepackage{dcolumn}
\usepackage{bm}
\usepackage{mathptmx}

\begin{document}

\title[]{Quantum Key Distribution over 100 km underwater optical fiber assisted by a \\ Fast-Gated Single-Photon Detector}


\author{Domenico Ribezzo}
\affiliation{Istituto Nazionale di Ottica del Consiglio Nazionale delle Ricerche (CNR-INO), 50125 Firenze, Italy}
\affiliation{Università degli Studi di Napoli Federico II, Napoli, Italy}

\author{Mujtaba Zahidy}
\affiliation{Centre of Excellence for Silicon Photonics for Optical Communications (SPOC), Department of Electrical and Photonics Engineering, Technical University of Denmark, Kgs. Lyngby, Denmark}

\author{Gianmarco Lemmi}
\affiliation{Istituto Nazionale di Ottica del Consiglio Nazionale delle Ricerche (CNR-INO), 50125 Firenze, Italy}
\affiliation{Università degli Studi di Napoli Federico II, Napoli, Italy}

\author{Antoine Petitjean}
\affiliation{Istituto Nazionale di Ottica del Consiglio Nazionale delle Ricerche (CNR-INO), 50125 Firenze, Italy}

\author{Claudia De Lazzari}
\affiliation{QTI S.r.l.,  50125, Firenze, Italy}

\author{Ilaria Vagniluca}
\affiliation{QTI S.r.l.,  50125, Firenze, Italy}

\author{Enrico Conca}
\affiliation{Dipartimento di Elettronica, Informazione e Bioingegneria, Politecnico di Milano, 20133 Milano, Italy}

\author{Alberto Tosi}
\affiliation{Dipartimento di Elettronica, Informazione e Bioingegneria, Politecnico di Milano, 20133 Milano, Italy}

\author{Tommaso Occhipinti}
\affiliation{QTI S.r.l.,  50125, Firenze, Italy}

\author{Leif K. Oxenl{\o}we}
\affiliation{Center for Silicon Photonics for Optical Communication (SPOC), Department of Photonics Engineering, Technical University of Denmark, Kgs. Lyngby, Denmark}

\author{André Xuereb}
\affiliation{Department of Physics, University of Malta, Msida MSD 2080, Malta}
\affiliation{Merqury Cybersecurity Limited, Malta}

\author{Davide Bacco}
\email{davide.bacco@unifi.it}
\affiliation{Department of Physics and Astronomy, University of Florence, 50019 Sesto Fiorentino, Italy}

\affiliation{QTI S.r.l.,  50125, Firenze, Italy}

\author{Alessandro Zavatta}
\email{alessandro.zavatta@ino.cnr.it}
\affiliation{Istituto Nazionale di Ottica del Consiglio Nazionale delle Ricerche (CNR-INO), 50125 Firenze, Italy}
\affiliation{QTI S.r.l.,  50125, Firenze, Italy}

\begin{abstract}
Nowadays Quantum Key Distribution represents the most mature quantum technology, and multiple countries as well as private institutions are building their quantum network. However, QKD devices are still far from representing a product within everyone's reach. Indeed, limitations in terms of compatibility with existing telecom infrastructure and limited performances in terms of secret key rate, using non-cryogenic detection systems, are still critical.
In this work, we implemented a quantum key distribution link between Sicily (Italy) and Malta utilizing two different Single-Photon Avalanche Diode (SPAD) detectors. The performances of a standard commercial SPAD have been compared with the results achieved with a new prototype of fast-gated System in a Package (SiP) SPAD; the SiP detector has shown to be able to accomplish a fourteen times higher key rate compared with the commercial device over the channel showing 20 dB of losses.
\end{abstract}
\maketitle

\section{Introduction}
Quantum Key Distribution (QKD), a method for exchanging symmetric cryptographic keys exploiting the laws of physics, is the most mature technology among the ones that appeared within the second quantum revolution \cite{bennett1984proceedings,lo2014secure,xu2020secure,scarani2009security}. Several experiments, both in physics laboratories and in field trial links, have shown QKD potential and readiness. Today, QKD links connecting cities among different continents are already a reality \cite{liao2017satellite,lu2022micius} and are employed in commercial applications as well as in governments. Nevertheless, many challenges still need to be faced in order to make QKD an everyday consumer technology. An important and very pragmatic example is the necessity to build QKD devices that are portable, scalable, and can guarantee a high key generation rate in long-distance links. In fact, today the current record in terms of key generation rate over a long-distance link has been achieved using Superconducting Nanowires Single-Photon Detectors (SNSPDs) \cite{liao2017satellite,lu2022micius,da2018record,neumann2022}, which present ultra-low dark count rates and high quantum efficiencies. The main drawback of this technology is its ultra-low operational temperature (below 4 K) which makes it difficult to integrate into deployable systems.

On the contrary, Single-Photon Avalanche Diodes (SPADs) working at room-temperature or at temperatures achievable with a compact cooling system offer high integrability in current telecommunication networks.

In this work, we realized a QKD link in the middle of the Mediterranean Sea, connecting Italy to Malta through a 100 km fiber-based underwater optical channel. The transmitter was located in a telecom center in the city of Pozzallo (Sicily, Italy), while the receiver was placed in the Melita Limited data center of Madliena (Malta). This link, which can be considered a new step in the frame of a European Quantum Network \cite{euroqci}, has been used to test a System in a Package (SiP) Indium Gallium Arsenide (InGaAs) SPAD \cite{signorelli2021low}. This detector features a dedicated fast-gated active quenching circuit that allows it to synchronize with a gate signal locked to the quantum states generation clock \cite{ruggeri2015integrated}. As a result, it is considerably less affected by dark counts and afterpulses. Moreover, we compared the SiP detector with a standard commercial InGaAs SPAD (ID221 by IDQuantique \cite{id221}). The new SPAD achieved a fourteen times higher key rate over the 20 dB-attenuation link with respect to the commercial device. We also investigated the behavior of the detector emulating a shorter link budget, showing that the SiP SPAD guarantees a high secret key rate up to 25 kbit/s at 3 dB channel loss .

Finally, we report a comparison of the detectors' performances in controlled laboratory conditions. We added to the comparison a second SiP detector, similar to the first one but with a larger sensor area, intended for free space applications.

\section{QKD Protocol}
The implemented protocol is the three-states efficient BB84 with time-bin encoding and one decoy method \cite{boaron2018secure,rusca2018finite,rusca2018security,hayashi2014security}. In this protocol, one basis is used for sharing the key, while the second basis is reserved for security checks. This choice allows to simplify the setup and to generate only one of the two eigenstates of the second mutually unbiased basis. The key generation basis is the computational $\mathbf{Z}$-basis, whose eigenstates, according to the time-bin encoding, are characterized by the emission time of a pulse into a time slot frame.
The eigenstates of the security check basis, $\mathbf{X}$-basis, are formed by the superposition of the $\mathbf{Z}$-basis with a relative phase (0 or $\pi$). It is worth pointing out that, even if the photon wave function is spread over two pulses, each state is supposed to contain no more than one photon; states with more photons (i.e. multi-photon states) introduce security issues and should be avoided. Unfortunately, multi-photon events cannot be totally suppressed, therefore, the decoy-state method has been introduced to overcome the vulnerabilities deriving from the lack of a real single-photon source \cite{hwang2003quantum,lo2005decoy}. In this method, randomly switching intensity levels helps to detect an eavesdropper that intercepts and re-sends only multi-photon states and blocks the rest and hence, cannot keep the photon number statistics stable. It has been proven that two different intensity levels are enough \cite{rusca2018finite}, a technique that is known as the one-decoy method.

For one-decoy 3-state BB84 protocol, in the finite-key regime, the key length $l$ is bound to \cite{boaron2018secure}:
\small
\begin{equation}
    l\leq s_{Z,0}^l+s_{Z,1}^l(1-H_2(\phi_Z^u))-\lambda_{EC}-6\log_2\left(\frac{19}{\epsilon_{sec}}\right)-\log_2\left(\frac{2}{\epsilon _{corr}}\right),
    \label{EQ::}
\end{equation}
\normalsize
with $s_{Z,0}^l$ and $s_{Z,1}^l$ being the lower bounds for the vacuum and the single-photon events, $\phi_Z^u$ the upper bound of the phase error rate, $\lambda_{EC}$ the number of disclosed bits in the error correction stage, $H_2(x)=-x\log_2(x)-(1-x)\log_2(1-x)$ the binary entropy and $\epsilon_{sec}=10^{-12}$ and $\epsilon_{corr}=10^{-12}$ the secrecy and correctness parameters. The $\epsilon$ parameters are defined as \cite{canale2014}:
\begin{equation*}
    \begin{split}
        P[S_A\neq S_B]  & < \epsilon_{corr}, \\
        \mathds{1}(S_A,S_B;Z,C) & < \epsilon_{sec},
    \end{split}
\end{equation*}
where $S_A$ and $S_B$ are Alice' and Bob's sifted keys, $P[x]$ the probability of $x$, $\mathds{1}(\cdot)$ a generic information measure, $Z$ is the eavesdropped sequence owned by a potential eavesdropper, and $C$ is a random variable representing the exchanged information. The second term denotes the probability $\epsilon_{sec}$ of having a stronger correlation between Alice's and Eve's strings than Alice's and Bob's ones. In the standard BB84, the phase error rate in the $\mathbf{Z}$-basis $\phi_Z$ corresponds to the bit error rate in the $\mathbf{X}$-basis $\delta_X$, however, since in this protocol Alice sends only one state in the $\mathbf{X}$-basis, $\phi_Z$ cannot be directly measured and it needs to be estimated from the $\mathbf{X}$-basis quantum bit error rate $\text{QBER}_X$ \cite{boaron2016detector}; it is connected to the visibility $\text{vis}_X$ of the receiver interferometer by $\text{QBER}_X=(1-\text{vis}_X)/2$.

\section{Experimental Setup}
\subsection{Network architecture and QKD devices}
The link is made by two 96 km long optical fibers deployed under the Mediterranean Sea and connecting Malta to Sicily; the same channel has already been employed for a demonstration of entanglement distribution in 2018 \cite{wengerowsky2018entanglement}. The fibers show an attenuation of around 20 and 21 dB, hence, we reserved the former for distributing the quantum states while the latter was used as a service channel (distribution of a synchronization signal, parameters estimation, etc.).

\begin{figure*}
    \centering
    \includegraphics[width=0.93\textwidth]{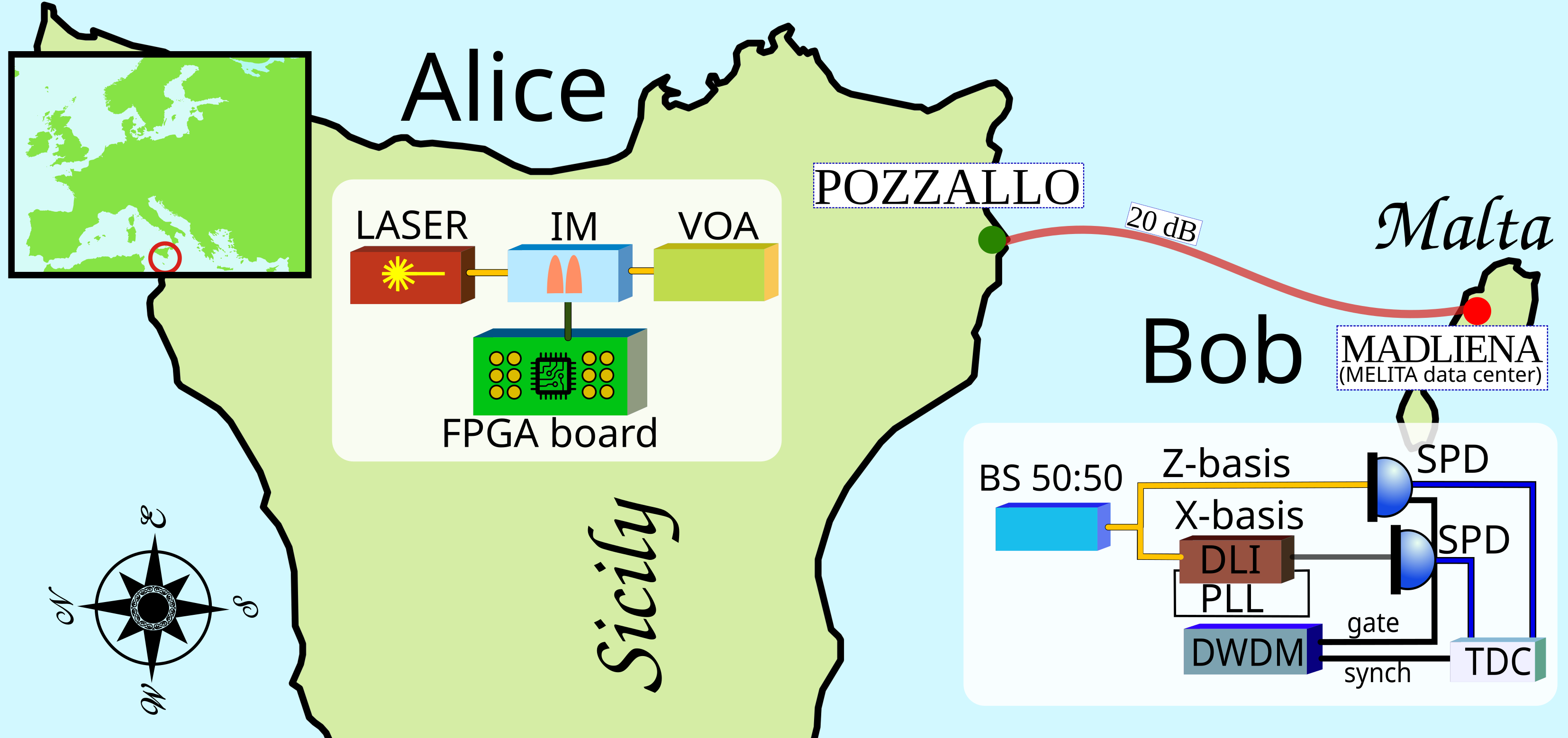}
    \caption{\textbf{Sketch of the setup.} Alice produces the states by carving (intensity modulator stage - IM) and attenuating (variable optical attenuation stage - VOA) a continuous wave laser. Bob makes the basis choice with a beam splitter (BS 50:50), then directly reads the arrival time of the photons (Z-basis) or makes an interferometric measurement with a delay line interferometer (DLI) for the X-basis. A second fiber is used to share a synchronization and a gate signal, multiplexed and then divided again by a dense wavelength division multiplexer (DWDM). The single-photon detectors are connected to a time-to-digit converter (TDC) that produces the timestamps to be elaborated by Bob's computer.}
    \label{fig:setup}
\end{figure*}

The experimental setup is illustrated in Fig. \ref{fig:setup}; the pulses encoding the states are generated by carving a continuous wave C-band laser with an intensity modulator controlled by a field programmable gate array (FPGA); after the carving stage, the pulses are attenuated down to single-photon level by a variable optical attenuator (VOA). More details about the transmitter device are reported in \cite{ribezzo2022}. The SiP SPAD can accept a gate trigger signal where the subsequent gating time (the ON and OFF times) of the detector can be set by the user.

The qubit generation rate has been fixed to 119 MHz for both detectors to acquire comparable data. However, the detector can accept up to a 150 MHz gate signal.

Alice and Bob select equal probabilities to generate and measure in the computational basis ($\mathbf{Z}$-basis), $P_{ZA}=P_{ZB}=0.5$; such choices for $P_{ZA}$ and $P_{ZB}$ are in accordance with a simulation model that takes into account the channel properties and the detection stage performances. 

On the service channel, two classical signals are shared between the two parts: a synchronization signal at 145 kHz and another signal at 119 MHz that is used as the gate signal for the detector. The mean numbers of photons per pulse are chosen such that they maximize the key rate in our simulation model, and are reported in tab. \ref{tab:results}.

After traveling through the underwater fiber channel, the photons arrive at the receiver setup; there they impinge on a 50:50 beam splitter, which acts as a passive basis choice. The $\mathbf{Z}$-basis output brings the photons directly to one Single-Photon Detector (SPD), while the $\mathbf{X}$-basis output lets the photons pass through a delay line interferometer (DLI) before reaching the detection part. The DLI is a Mach-Zehnder interferometer with one arm  800 ps longer than the other, so that the two pulses characterizing the wave-function states in the $\mathbf{X}$-basis overlap and their relative phase can be measured. 
The interferometer is stabilized by a phase-lock loop (PLL) which adjusts a phase shifter to compensate for phase fluctuations. The feedback for the loop is provided by sending a weak classical laser, counter-propagating with respect to the quantum signal, and monitoring its phase fluctuation.
Finally, the synchronization and the gate signals traveling in the service channel are demultiplexed and 
sent to the corresponding modules.

\begin{table}[]
    \centering
	    \begin{tabular}{c|c|c}
         & \textbf{SiP PoliMi} & \textbf{ID221} \\
         \hline
         $\tau_{off}$ ($\mu$s) & 1 & 20\\
         $\textbf{r}_{\text{DC}}$ (kHz) & 10.8 & 2.5 \\
         $n_Z$ & \multicolumn{2}{c}{$10^9$} \\ 
         $p_{Z,A}$ & \multicolumn{2}{c}{50\%}\\
         $p_{Z_,B}$ & \multicolumn{2}{c}{50\%}\\
         $\nu_{rep}$ (MHz) & \multicolumn{2}{c}{119}\\
         $\epsilon_{sec}$ & \multicolumn{2}{c}{$10^{-12}$} \\
         $\epsilon_{corr}$ & \multicolumn{2}{c}{$10^{-12}$}\\
         $\tau_Z$ (dB) & \multicolumn{2}{c}{1} \\
         $\tau_X$ (dB) & \multicolumn{2}{c}{3}
    \end{tabular}
\caption{\textbf{Setup parameters.} $\tau_{off}$ is the hold-off time of the detectors, $\textbf{r}_{DC}$ the dark count rate, $n_Z$ is the block size, $p_{Z,A}$ and $p_{Z,B}$ the probabilities of choosing the Z basis for Alice and Bob respectively, $\nu_{rep}$ is the repetition rate, $\epsilon_{sec}$ and $\epsilon_{corr}$ are the security and correctness parameters, $\tau_Z$ and $\tau_X$ are the losses of Bob for the Z and X basis. SiP PoliMi detector is the SiP detector with 10 $\mu$m diameter sensitive area.}
    
    \label{tab:parameters}
\end{table}

\subsection{Detecting stage}
The employed research-product SiP detector is a state-of-the-art InGaAs/InP SPAD developed at Politecnico di Milano (PoliMi) and designed to operate with low dark count rate, competitive photon detection efficiency, and contained timing jitter. The primary feature of this detector is its time-gating capability. A conventional gated circuits often uses a simple passive quenching circuit, which cannot be gated at high frequency and requires a long dead time to limit the afterpulsing effect (APE). APE happens when carriers generated in an avalanche are trapped, and after a certain time (up to a few microseconds for InGaAs/InP SPADs operating at 220 - 240 K), are randomly released, resulting in a secondary avalanche without any real photon impinging on the SPAD. By implementing a fast active quenching circuit in place of a simple passive one, the after-pulses are strongly reduced. The tested detector implements a newly developed circuit able to fast-gate the detector at frequencies up to 150 MHz, with ON-time as short as few hundreds of picoseconds. When a photon is detected, this circuit enforces a hold-off time to the SPAD by skipping a programmable number of gate periods, resulting in a suppression of the after-pulses impact \cite{ruggeri2015integrated}. The photosensitive area of the SPAD has a diameter of around 10 $\mu$m, making it the perfect choice for fiber-based applications. A second detector with identical characteristics except for a bigger sensitive area (25 $\mu$m) has been tested utilizing the same optical setup. The paper \cite{tosi2012ingaas} reports a detailed description of the detector and an accurate characterization of its specifications in laboratory conditions.

\section{Results}
\subsection{Field trial}
The described setup has been utilized for establishing a QKD protocol from Sicily to Malta. We performed the experiment and data acquisition with both the ID221 detector from IDQuantique and the described SiP PoliMi SPAD.
\begin{table}[]
\centering
    \begin{tabular}{c|c|c|c|c|c}
         & 3 dB & 5 dB & 10 dB & 15 dB & 20 dB \\
         \hline
         \multicolumn{6}{c}{SPAD by Polimi}\\
         \hline
         $\mu_1$ & 0.36 & 0.41 & 0.46 & 0.46 & 0.41\\
         $\mu_2$ & 0.16 & 0.16 & 0.16 & 0.16 & 0.16\\
         $\mu_3$ & \multicolumn{5}{c}{0}\\
         $\epsilon_{Z} (\%)$ & 0.7 & 0.8 & 1.1 & 1.8 & 4.6 \\
         $\epsilon_{X} (\%)$ & 2.8 & 3.0 & 3.1 & 3.4 & 6.4\\
        SKR (kbps) & 24.65 & 21.75 & 13.10 & 5.80 & 1.50\\ 
       \hline     
		\hline
         \multicolumn{6}{c}{ID221 SPAD}\\
         \hline
         $\mu_1$ & 0.21 & 0.31 & 0.31 & 0.36 & 0.41\\
         $\mu_2$ & 0.06 & 0.11 & 0.11 & 0.16 & 0.16\\
         $\mu_3$ & \multicolumn{5}{c}{0}\\
         5ttt$\epsilon_{Z} (\%)$ & 4.4 & 4.4 & 5.0 & 6.0 & 9.3\\
         $\epsilon_{X} (\%)$ & 4 & 2.9 & 3.2 & 4.0 & 7.2\\
		SKR (kbps) & 3.25 & 3.05 & 2.10 & 1.05 & 0.11\\          
    \end{tabular}
    \caption{\textbf{Chosen parameters and measured values:} $\mu_1$, $\mu_2$ and $\mu_3$ are the numbers of photons per pulse according to the decoy method, $\epsilon_Z$ and $\epsilon_X$ are the qubit error rate in the two bases and finally, SKR is the secure key rate. The probabilities of choosing each $\mu$ have been chosen such that it maximizes the key rate.}
    \label{tab:results}
\end{table}
For the ID221, a hold-off time of $\tau_{\text{off}} = 20\mu$s has been set in order to keep the after-pulses within manageable values. The hold-off feature keeps the SPAD turned off for $\tau_{\text{off}}$ after each detection event to empty the active area from possibly trapped carriers. For the fast-gated detector, we have been able to set $\tau_{\text{off}} = 1\mu$s thanks to the limited after-pulse probability.

With the hold-off time set to 1 $\mu$s, the SiP PoliMi detector shows a higher dark count rate compared to the commercial SPAD (10.8 kHz vs 2.5 KHz). However, the maximum count rate $CR_{\max}=1/\tau_{\text{off}}$ allows to detect a higher rate of events than the commercial SPAD. It should be noted that a high $\tau_{\text{off}}$ also limits the SPAD performance by reducing its saturation threshold, resulting in lower detection efficiency. The low $\tau_{\text{off}}$ setting allows for avoiding such conditions in the SiP PoliMi detector.

In comparison, ID221 shows a dark count rate of around 200 kHz for $\tau_{\text{off}} = 1\mu$s. A specially designed quenching circuit that manages the fast-gate signal allows considerably improved performances for the SiP PoliMi SPAD at low $\tau_{\text{off}}$.

Successively, to evaluate the performance of the detector at different channel losses, we repeat the experiment on shorter segments of the channel. To simulate that, we gradually compensated for the losses encountered by photons traveling at different channel lengths by increasing the input power. This is equivalent to placing Alice's transmitter in the corresponding loss-compensated location on the link.

\begin{figure}
    \centering
    \includegraphics[width=0.98\columnwidth]{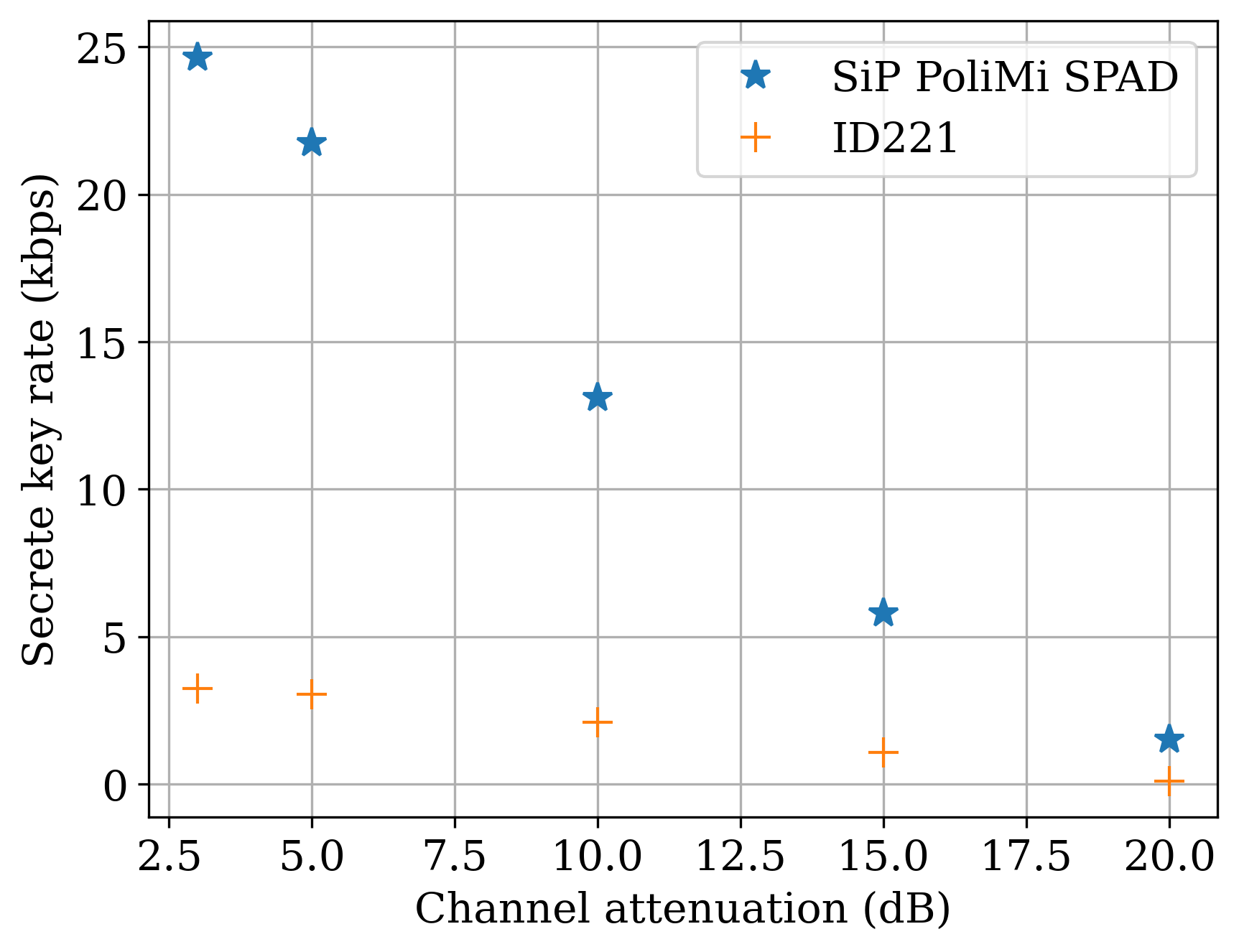}
    \caption{\textbf{Key rate results (field trial).} The plot shows the secure key rates achieved by the commercial detector (ID221) and the tested SiP detector fabricated by the research group of Politecnico di Milano (SiP Polimi SPAD). Thanks to the fast-quenching circuit applying the gate signal, the SiP PoliMi SPAD outperforms the commercial device by a factor of seven in terms of key rate for small attenuation link, and up to fourteen times when the entire channel is considered (20 dB).}
    \label{fig:comparison_malta}
\end{figure}

The achieved key rates are reported in Tab. \ref{tab:results} and are shown in Fig. \ref{fig:comparison_malta}.
\subsection{Laboratory test}
\begin{figure}[htbp]
    \centering
    \includegraphics[width=0.98\columnwidth]{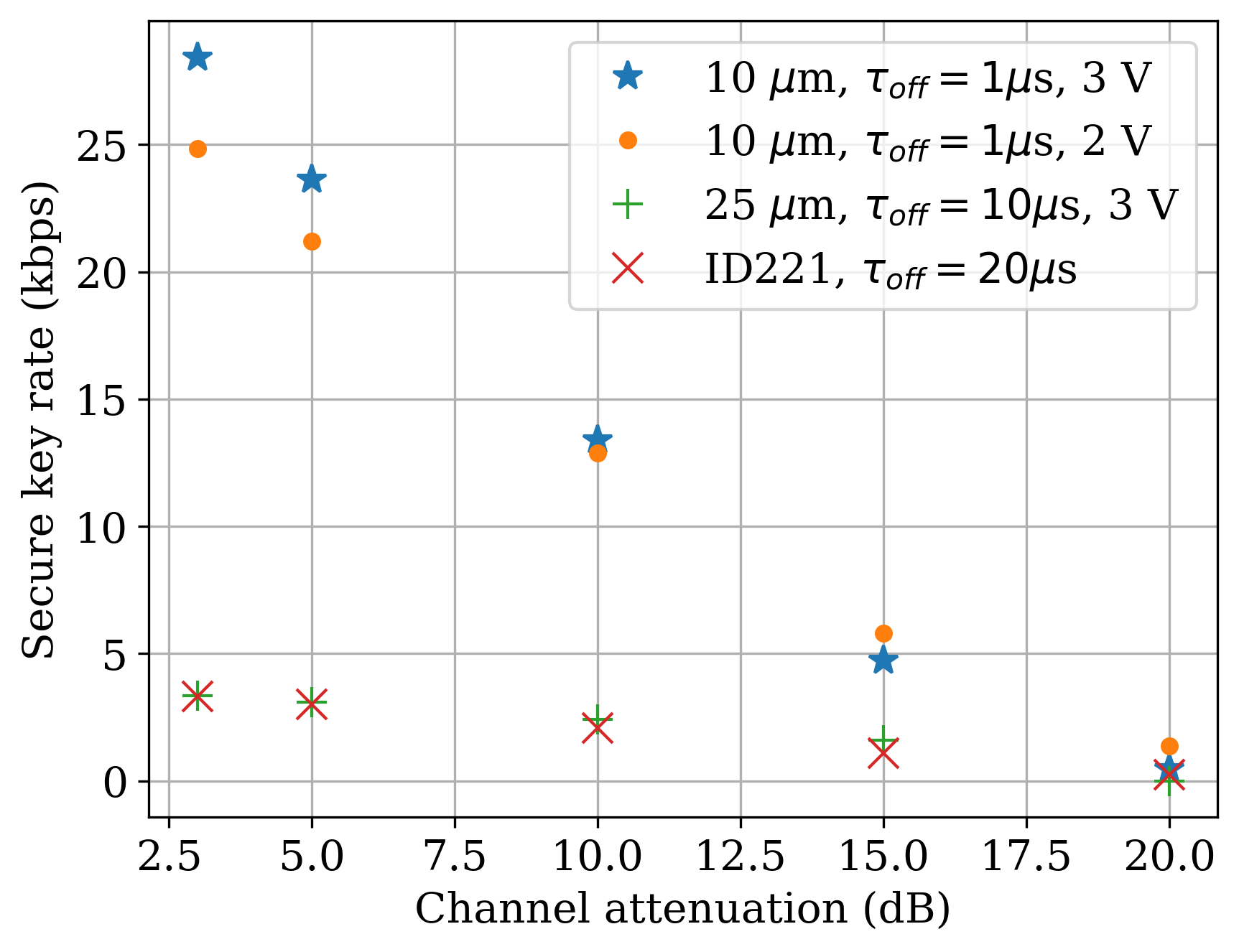}
    \caption{Secrete key rate extracted in laboratory conditions with the different detectors for different channel attenuation values. 10 $\mu$m and 25 $\mu$m are the two SiP PoliMi SPAD and the value is referred to the relative photosensitive active area diameter. The selected detector settings are reported in the legend (hold-off time, excess bias voltage).}
    \label{fig:comparison_lab}
\end{figure}

The detector has successively been tested in controlled laboratory conditions. The second SiP detector with a sensitive area diameter of 25 $\mu$m has been added to the comparison. Since a bigger sensitive area entails a higher dark count rate, a hold-off time of 10 $\mu$s has been preferred for this detector. We performed the test for different channel losses introduced by a tunable attenuator between Alice's and Bob's setups. The results are reported in Tab. \ref{tab:lab_comp} and Fig. \ref{fig:comparison_lab}. The second SiP PoliMi detector does not show a significant improvement over the commercial SPAD. We observed a small-scale increase in SKR up to 15 dB of channel loss, however, due to excess noise and low signal-to-noise ratio (SNR), SKR falls to zero at higher channel losses. Finally, the SiP detector with the smaller active area (10 $\mu$m diameter) was tested under different excess bias voltages. While increasing excess bias improves the detector's efficiency, it also increases the afterpulsing effect and the dark count rate. The test results show an improvement in low channel losses, while the performance dropped with excess bias voltage increase due to the reduction of SNR. 

\begin{table}[]
    \centering
    \begin{tabular}{c|c|c|c|c}
    \textbf{3 dB} & \textbf{5 dB} & \textbf{10 dB} & \textbf{15 dB} & \textbf{20 dB}\\
    \hline
    \hline
    \multicolumn{5}{c}{10 $\mu$m SPAD, 1 $\mu$s, 2V (kbps)}\\
    \hline
    24.83 & 21.19 & 12.89 & 5.80 & 1.38\\
    \hline
    \hline
    \multicolumn{5}{c}{10 $\mu$m SPAD, 1 $\mu$s, 3V (kbps)}\\
    \hline
    28.42 & 23.62 & 13.39 & 4.72 & 0.47\\
    \hline
    \hline
    \multicolumn{5}{c}{25 $\mu$m SPAD, 10 $\mu$s, 3V (kbps)}\\
    \hline
    3.34 & 3.10 & 2.42 & 1.61 & 0\\
     \hline
    \hline
    \multicolumn{5}{c}{IDQ, 20 $\mu$s, 20\% eff. (kbps)}\\
    \hline
    3.32 & 3.00 & 2.07 & 1.10 & 0.15\\
    \end{tabular}
    \caption{Secure key rate extracted with the two SiP PoliMi detectors in laboratory conditions. The detector with 10 $\mu$m sensitive area has been set with a hold-off time of 1 $\mu$s and has been tested for 2V and 3V of excess bias. For the detector with a sensor area of 25 $\mu$m, the set parameters are 10 $\mu$s of hold-off time and 3V of excess bias. The ID221 has been kept on 20 $\mu$s of hold-off time and 20\% of detection efficiency.}
    \label{tab:lab_comp}
\end{table}

\section{Discussion}
Boosting the key rate on long-distance links is the priority for the widespread deployment of QKD technology.  Many new protocols are being experimented and are showing their potentialities: twin field QKD \cite{lucamarini2018overcoming} and high dimensional protocols \cite{bechmann2000quantum,vagniluca2020efficient,zahidy20224,da2021path,bacco2021characterization} are just few examples that go in this direction. Simultaneously, with the appearance of SNSPDs \cite{natarajan2012superconducting,dauler2014review} and photon number resolving detectors \cite{divochiy2008superconducting,provaznik2020benchmarking} much attention has been paid to the detection stage. Although InGaAs/InP SPADs established their place as the most common technology for single-photons detection in C-band because of their portability and cost-effectiveness, compared to newer technologies, they present limited performance in terms of quantum efficiency, dead-time, and timing jitter. In addition, they are considerably more affected by dark counts and afterpulsing phenomenon.

In this paper, we enabled quantum communications between two European countries. Although several works already demonstrated a limited European quantum network, a full-scale deployment faces many open challenges regarding range, cost, etc \cite{ribezzo2022}. This work, introducing a more cost-effective approach, represents an additional step toward a European quantum infrastructure.

We demonstrated that in a real QKD scenario, without an important technological replacement and only improving the detection stage, an improvement of up to a factor of 14 in terms of key rate is achievable thanks to an advanced sensor design and an active quenching circuit implementation.

The advanced innovative built-in quenching circuit together with the adaptive gating technique allows for increasing the detection rate as well as reducing the effect of afterpulsing by minimizing the ON time of the detector to the expected optical pulse width. Thanks to the fast quenching, a gating of up to 150 MHz is achievable, which contributes to the final key rate in significant amounts. In comparison, in the old technology, a long hold-off time was necessary to overcome the effect of afterpulsing, which in turn reduces the detection rate. Besides, the two SiP SPADs have been engineered and designed in order to show state-of-the-art performances in terms of intrinsic dark counts, timing jitter, and detection efficiency.

This work also provides a comparison of performance with the sensitive area dimensions. The second SiP PoliMi detector featuring a 25 $\mu$m diameter active area and similar circuitry shows greater susceptibility to dark counts and afterpulsing. In comparison, the new detector produces results slightly better than the commercial detector. It should be noted that the 25 $\mu$m detector has been designed and intended for free-space applications where a larger sensitive area is desiderable.

Our demonstration proves the effectiveness of the newly introduced detector technology in reducing cost per secure bit and increasing the final key generation rate, and will help to make QKD a more user-accessible technology.

\section{Acknowledgements}
This work was partially supported by the Center of Excellence SPOC (ref DNRF123), Innovations fonden project Fire-Q (No. 9090-00031B), the NATO Science for Peace and Security program (Grant No. G5485, project SEQUEL), the programme Rita Levi Montalcini QOMUNE (PGR19GKW5T), the EraNET Cofund Initiatives QuantERA within the European Union’s Horizon 2020 research and innovation program grant agreement No.731473 (project SQUARE), the Project EQUO (European QUantum ecOsystems) which is funded by the European Commission in the Digital Europe Programme under the grant agreement No 101091561, the Project SERICS (PE00000014) under the MUR National Recovery and Resilience Plan funded by the European Union - NextGenerationEU and the Project QuONTENT under the “Progetti di Ricerca@CNR” program funded by the Consiglio Nazionale delle Ricerche (CNR).

\bibliography{mybib}

\begin{thebibliography}{33}%
\makeatletter
\providecommand \@ifxundefined [1]{%
 \@ifx{#1\undefined}
}%
\providecommand \@ifnum [1]{%
 \ifnum #1\expandafter \@firstoftwo
 \else \expandafter \@secondoftwo
 \fi
}%
\providecommand \@ifx [1]{%
 \ifx #1\expandafter \@firstoftwo
 \else \expandafter \@secondoftwo
 \fi
}%
\providecommand \natexlab [1]{#1}%
\providecommand \enquote  [1]{``#1''}%
\providecommand \bibnamefont  [1]{#1}%
\providecommand \bibfnamefont [1]{#1}%
\providecommand \citenamefont [1]{#1}%
\providecommand \href@noop [0]{\@secondoftwo}%
\providecommand \href [0]{\begingroup \@sanitize@url \@href}%
\providecommand \@href[1]{\@@startlink{#1}\@@href}%
\providecommand \@@href[1]{\endgroup#1\@@endlink}%
\providecommand \@sanitize@url [0]{\catcode `\\12\catcode `\$12\catcode
  `\&12\catcode `\#12\catcode `\^12\catcode `\_12\catcode `\%12\relax}%
\providecommand \@@startlink[1]{}%
\providecommand \@@endlink[0]{}%
\providecommand \url  [0]{\begingroup\@sanitize@url \@url }%
\providecommand \@url [1]{\endgroup\@href {#1}{\urlprefix }}%
\providecommand \urlprefix  [0]{URL }%
\providecommand \Eprint [0]{\href }%
\providecommand \doibase [0]{https://doi.org/}%
\providecommand \selectlanguage [0]{\@gobble}%
\providecommand \bibinfo  [0]{\@secondoftwo}%
\providecommand \bibfield  [0]{\@secondoftwo}%
\providecommand \translation [1]{[#1]}%
\providecommand \BibitemOpen [0]{}%
\providecommand \bibitemStop [0]{}%
\providecommand \bibitemNoStop [0]{.\EOS\space}%
\providecommand \EOS [0]{\spacefactor3000\relax}%
\providecommand \BibitemShut  [1]{\csname bibitem#1\endcsname}%
\let\auto@bib@innerbib\@empty
\bibitem [{\citenamefont {Bennett}\ and\ \citenamefont
  {Brassard}(2014)}]{bennett1984proceedings}%
  \BibitemOpen
  \bibfield  {author} {\bibinfo {author} {\bibfnamefont {C.~H.}\ \bibnamefont
  {Bennett}}\ and\ \bibinfo {author} {\bibfnamefont {G.}~\bibnamefont
  {Brassard}},\ }\bibfield  {title} {\bibinfo {title} {Quantum cryptography:
  Public key distribution and coin tossing},\ }\href
  {https://doi.org/https://doi.org/10.1016/j.tcs.2014.05.025} {\bibfield
  {journal} {\bibinfo  {journal} {Theoretical Computer Science}\ }\textbf
  {\bibinfo {volume} {560}},\ \bibinfo {pages} {7} (\bibinfo {year} {2014})},\
  \bibinfo {note} {theoretical Aspects of Quantum Cryptography – celebrating
  30 years of BB84}\BibitemShut {NoStop}%
\bibitem [{\citenamefont {Lo}\ \emph {et~al.}(2014)\citenamefont {Lo},
  \citenamefont {Curty},\ and\ \citenamefont {Tamaki}}]{lo2014secure}%
  \BibitemOpen
  \bibfield  {author} {\bibinfo {author} {\bibfnamefont {H.-K.}\ \bibnamefont
  {Lo}}, \bibinfo {author} {\bibfnamefont {M.}~\bibnamefont {Curty}},\ and\
  \bibinfo {author} {\bibfnamefont {K.}~\bibnamefont {Tamaki}},\ }\bibfield
  {title} {\bibinfo {title} {Secure quantum key distribution},\ }\href
  {https://doi.org/10.1038/nphoton.2014.149} {\bibfield  {journal} {\bibinfo
  {journal} {Nature Photonics}\ }\textbf {\bibinfo {volume} {8}},\ \bibinfo
  {pages} {595} (\bibinfo {year} {2014})}\BibitemShut {NoStop}%
\bibitem [{\citenamefont {Xu}\ \emph {et~al.}(2020)\citenamefont {Xu},
  \citenamefont {Ma}, \citenamefont {Zhang}, \citenamefont {Lo},\ and\
  \citenamefont {Pan}}]{xu2020secure}%
  \BibitemOpen
  \bibfield  {author} {\bibinfo {author} {\bibfnamefont {F.}~\bibnamefont
  {Xu}}, \bibinfo {author} {\bibfnamefont {X.}~\bibnamefont {Ma}}, \bibinfo
  {author} {\bibfnamefont {Q.}~\bibnamefont {Zhang}}, \bibinfo {author}
  {\bibfnamefont {H.-K.}\ \bibnamefont {Lo}},\ and\ \bibinfo {author}
  {\bibfnamefont {J.-W.}\ \bibnamefont {Pan}},\ }\bibfield  {title} {\bibinfo
  {title} {Secure quantum key distribution with realistic devices},\ }\href
  {https://doi.org/10.1103/RevModPhys.92.025002} {\bibfield  {journal}
  {\bibinfo  {journal} {Rev. Mod. Phys.}\ }\textbf {\bibinfo {volume} {92}},\
  \bibinfo {pages} {025002} (\bibinfo {year} {2020})}\BibitemShut {NoStop}%
\bibitem [{\citenamefont {Scarani}\ \emph {et~al.}(2009)\citenamefont
  {Scarani}, \citenamefont {Bechmann-Pasquinucci}, \citenamefont {Cerf},
  \citenamefont {Du\ifmmode~\check{s}\else \v{s}\fi{}ek}, \citenamefont
  {L\"utkenhaus},\ and\ \citenamefont {Peev}}]{scarani2009security}%
  \BibitemOpen
  \bibfield  {author} {\bibinfo {author} {\bibfnamefont {V.}~\bibnamefont
  {Scarani}}, \bibinfo {author} {\bibfnamefont {H.}~\bibnamefont
  {Bechmann-Pasquinucci}}, \bibinfo {author} {\bibfnamefont {N.~J.}\
  \bibnamefont {Cerf}}, \bibinfo {author} {\bibfnamefont {M.}~\bibnamefont
  {Du\ifmmode~\check{s}\else \v{s}\fi{}ek}}, \bibinfo {author} {\bibfnamefont
  {N.}~\bibnamefont {L\"utkenhaus}},\ and\ \bibinfo {author} {\bibfnamefont
  {M.}~\bibnamefont {Peev}},\ }\bibfield  {title} {\bibinfo {title} {The
  security of practical quantum key distribution},\ }\href
  {https://doi.org/10.1103/RevModPhys.81.1301} {\bibfield  {journal} {\bibinfo
  {journal} {Rev. Mod. Phys.}\ }\textbf {\bibinfo {volume} {81}},\ \bibinfo
  {pages} {1301} (\bibinfo {year} {2009})}\BibitemShut {NoStop}%
\bibitem [{\citenamefont {Liao}\ \emph {et~al.}(2017)\citenamefont {Liao},
  \citenamefont {Cai}, \citenamefont {Liu}, \citenamefont {Zhang},
  \citenamefont {Li}, \citenamefont {Ren}, \citenamefont {Yin}, \citenamefont
  {Shen}, \citenamefont {Cao}, \citenamefont {Li}, \citenamefont {Li},
  \citenamefont {Chen}, \citenamefont {Sun}, \citenamefont {Jia}, \citenamefont
  {Wu}, \citenamefont {Jiang}, \citenamefont {Wang}, \citenamefont {Huang},
  \citenamefont {Wang}, \citenamefont {Zhou}, \citenamefont {Deng},
  \citenamefont {Xi}, \citenamefont {Ma}, \citenamefont {Hu}, \citenamefont
  {Zhang}, \citenamefont {Chen}, \citenamefont {Liu}, \citenamefont {Wang},
  \citenamefont {Zhu}, \citenamefont {Lu}, \citenamefont {Shu}, \citenamefont
  {Peng}, \citenamefont {Wang},\ and\ \citenamefont {Pan}}]{liao2017satellite}%
  \BibitemOpen
  \bibfield  {author} {\bibinfo {author} {\bibfnamefont {S.-K.}\ \bibnamefont
  {Liao}}, \bibinfo {author} {\bibfnamefont {W.-Q.}\ \bibnamefont {Cai}},
  \bibinfo {author} {\bibfnamefont {W.-Y.}\ \bibnamefont {Liu}}, \bibinfo
  {author} {\bibfnamefont {L.}~\bibnamefont {Zhang}}, \bibinfo {author}
  {\bibfnamefont {Y.}~\bibnamefont {Li}}, \bibinfo {author} {\bibfnamefont
  {J.-G.}\ \bibnamefont {Ren}}, \bibinfo {author} {\bibfnamefont
  {J.}~\bibnamefont {Yin}}, \bibinfo {author} {\bibfnamefont {Q.}~\bibnamefont
  {Shen}}, \bibinfo {author} {\bibfnamefont {Y.}~\bibnamefont {Cao}}, \bibinfo
  {author} {\bibfnamefont {Z.-P.}\ \bibnamefont {Li}}, \bibinfo {author}
  {\bibfnamefont {F.-Z.}\ \bibnamefont {Li}}, \bibinfo {author} {\bibfnamefont
  {X.-W.}\ \bibnamefont {Chen}}, \bibinfo {author} {\bibfnamefont {L.-H.}\
  \bibnamefont {Sun}}, \bibinfo {author} {\bibfnamefont {J.-J.}\ \bibnamefont
  {Jia}}, \bibinfo {author} {\bibfnamefont {J.-C.}\ \bibnamefont {Wu}},
  \bibinfo {author} {\bibfnamefont {X.-J.}\ \bibnamefont {Jiang}}, \bibinfo
  {author} {\bibfnamefont {J.-F.}\ \bibnamefont {Wang}}, \bibinfo {author}
  {\bibfnamefont {Y.-M.}\ \bibnamefont {Huang}}, \bibinfo {author}
  {\bibfnamefont {Q.}~\bibnamefont {Wang}}, \bibinfo {author} {\bibfnamefont
  {Y.-L.}\ \bibnamefont {Zhou}}, \bibinfo {author} {\bibfnamefont
  {L.}~\bibnamefont {Deng}}, \bibinfo {author} {\bibfnamefont {T.}~\bibnamefont
  {Xi}}, \bibinfo {author} {\bibfnamefont {L.}~\bibnamefont {Ma}}, \bibinfo
  {author} {\bibfnamefont {T.}~\bibnamefont {Hu}}, \bibinfo {author}
  {\bibfnamefont {Q.}~\bibnamefont {Zhang}}, \bibinfo {author} {\bibfnamefont
  {Y.-A.}\ \bibnamefont {Chen}}, \bibinfo {author} {\bibfnamefont {N.-L.}\
  \bibnamefont {Liu}}, \bibinfo {author} {\bibfnamefont {X.-B.}\ \bibnamefont
  {Wang}}, \bibinfo {author} {\bibfnamefont {Z.-C.}\ \bibnamefont {Zhu}},
  \bibinfo {author} {\bibfnamefont {C.-Y.}\ \bibnamefont {Lu}}, \bibinfo
  {author} {\bibfnamefont {R.}~\bibnamefont {Shu}}, \bibinfo {author}
  {\bibfnamefont {C.-Z.}\ \bibnamefont {Peng}}, \bibinfo {author}
  {\bibfnamefont {J.-Y.}\ \bibnamefont {Wang}},\ and\ \bibinfo {author}
  {\bibfnamefont {J.-W.}\ \bibnamefont {Pan}},\ }\bibfield  {title} {\bibinfo
  {title} {Satellite-to-ground quantum key distribution},\ }\href
  {https://doi.org/10.1038/nature23655} {\bibfield  {journal} {\bibinfo
  {journal} {Nature}\ }\textbf {\bibinfo {volume} {549}},\ \bibinfo {pages}
  {43} (\bibinfo {year} {2017})}\BibitemShut {NoStop}%
\bibitem [{\citenamefont {Lu}\ \emph {et~al.}(2022)\citenamefont {Lu},
  \citenamefont {Cao}, \citenamefont {Peng},\ and\ \citenamefont
  {Pan}}]{lu2022micius}%
  \BibitemOpen
  \bibfield  {author} {\bibinfo {author} {\bibfnamefont {C.-Y.}\ \bibnamefont
  {Lu}}, \bibinfo {author} {\bibfnamefont {Y.}~\bibnamefont {Cao}}, \bibinfo
  {author} {\bibfnamefont {C.-Z.}\ \bibnamefont {Peng}},\ and\ \bibinfo
  {author} {\bibfnamefont {J.-W.}\ \bibnamefont {Pan}},\ }\bibfield  {title}
  {\bibinfo {title} {Micius quantum experiments in space},\ }\href
  {https://doi.org/10.1103/RevModPhys.94.035001} {\bibfield  {journal}
  {\bibinfo  {journal} {Rev. Mod. Phys.}\ }\textbf {\bibinfo {volume} {94}},\
  \bibinfo {pages} {035001} (\bibinfo {year} {2022})}\BibitemShut {NoStop}%
\bibitem [{\citenamefont {Lio}\ \emph {et~al.}(2018)\citenamefont {Lio},
  \citenamefont {Bacco}, \citenamefont {Cozzolino}, \citenamefont {Ros},
  \citenamefont {Guo}, \citenamefont {Ding}, \citenamefont {Sasaki},
  \citenamefont {Aikawa}, \citenamefont {Miki}, \citenamefont {Terai},
  \citenamefont {Yamashita}, \citenamefont {Neergaard-Nielsen}, \citenamefont
  {Galili}, \citenamefont {Rottwitt}, \citenamefont {Andersen}, \citenamefont
  {Oxenl{\o}we},\ and\ \citenamefont {Morioka}}]{da2018record}%
  \BibitemOpen
  \bibfield  {author} {\bibinfo {author} {\bibfnamefont {B.~D.}\ \bibnamefont
  {Lio}}, \bibinfo {author} {\bibfnamefont {D.}~\bibnamefont {Bacco}}, \bibinfo
  {author} {\bibfnamefont {D.}~\bibnamefont {Cozzolino}}, \bibinfo {author}
  {\bibfnamefont {F.~D.}\ \bibnamefont {Ros}}, \bibinfo {author} {\bibfnamefont
  {X.}~\bibnamefont {Guo}}, \bibinfo {author} {\bibfnamefont {Y.}~\bibnamefont
  {Ding}}, \bibinfo {author} {\bibfnamefont {Y.}~\bibnamefont {Sasaki}},
  \bibinfo {author} {\bibfnamefont {K.}~\bibnamefont {Aikawa}}, \bibinfo
  {author} {\bibfnamefont {S.}~\bibnamefont {Miki}}, \bibinfo {author}
  {\bibfnamefont {H.}~\bibnamefont {Terai}}, \bibinfo {author} {\bibfnamefont
  {T.}~\bibnamefont {Yamashita}}, \bibinfo {author} {\bibfnamefont {J.~S.}\
  \bibnamefont {Neergaard-Nielsen}}, \bibinfo {author} {\bibfnamefont
  {M.}~\bibnamefont {Galili}}, \bibinfo {author} {\bibfnamefont
  {K.}~\bibnamefont {Rottwitt}}, \bibinfo {author} {\bibfnamefont {U.~L.}\
  \bibnamefont {Andersen}}, \bibinfo {author} {\bibfnamefont {L.~K.}\
  \bibnamefont {Oxenl{\o}we}},\ and\ \bibinfo {author} {\bibfnamefont
  {T.}~\bibnamefont {Morioka}},\ }\bibfield  {title} {\bibinfo {title}
  {Record-high secret key rate for joint classical and quantum transmission
  over a 37-core fiber},\ }in\ \href
  {https://doi.org/10.1109/IPCon.2018.8527341} {\emph {\bibinfo {booktitle}
  {2018 IEEE Photonics Conference (IPC)}}}\ (\bibinfo {year} {2018})\ pp.\
  \bibinfo {pages} {1--2}\BibitemShut {NoStop}%
\bibitem [{\citenamefont {Neumann}\ \emph {et~al.}(2022)\citenamefont
  {Neumann}, \citenamefont {Buchner}, \citenamefont {Bulla}, \citenamefont
  {Bohmann},\ and\ \citenamefont {Ursin}}]{neumann2022}%
  \BibitemOpen
  \bibfield  {author} {\bibinfo {author} {\bibfnamefont {S.~P.}\ \bibnamefont
  {Neumann}}, \bibinfo {author} {\bibfnamefont {A.}~\bibnamefont {Buchner}},
  \bibinfo {author} {\bibfnamefont {L.}~\bibnamefont {Bulla}}, \bibinfo
  {author} {\bibfnamefont {M.}~\bibnamefont {Bohmann}},\ and\ \bibinfo {author}
  {\bibfnamefont {R.}~\bibnamefont {Ursin}},\ }\bibfield  {title} {\bibinfo
  {title} {Continuous entanglement distribution over a transnational
  248{\thinspace}km fiber link},\ }\href
  {https://doi.org/10.1038/s41467-022-33919-0} {\bibfield  {journal} {\bibinfo
  {journal} {Nature Communications}\ }\textbf {\bibinfo {volume} {13}},\
  \bibinfo {pages} {6134} (\bibinfo {year} {2022})}\BibitemShut {NoStop}%
\bibitem [{\citenamefont {{EuroQCI}}(2017)}]{euroqci}%
  \BibitemOpen
  \bibfield  {author} {\bibinfo {author} {\bibnamefont {{EuroQCI}}},\
  }\href@noop {} {\bibinfo {title} {European quantum communication
  infrastructure (euroqci) initiative}} (\bibinfo {year} {2017}),\ \bibinfo
  {note}
  {https://digital-strategy.ec.europa.eu/en/policies/european-quantum-communication-infrastructure-euroqci}\BibitemShut
  {NoStop}%
\bibitem [{\citenamefont {Signorelli}\ \emph {et~al.}(2022)\citenamefont
  {Signorelli}, \citenamefont {Telesca}, \citenamefont {Conca}, \citenamefont
  {Frera}, \citenamefont {Ruggeri}, \citenamefont {Giudice},\ and\
  \citenamefont {Tosi}}]{signorelli2021low}%
  \BibitemOpen
  \bibfield  {author} {\bibinfo {author} {\bibfnamefont {F.}~\bibnamefont
  {Signorelli}}, \bibinfo {author} {\bibfnamefont {F.}~\bibnamefont {Telesca}},
  \bibinfo {author} {\bibfnamefont {E.}~\bibnamefont {Conca}}, \bibinfo
  {author} {\bibfnamefont {A.~D.}\ \bibnamefont {Frera}}, \bibinfo {author}
  {\bibfnamefont {A.}~\bibnamefont {Ruggeri}}, \bibinfo {author} {\bibfnamefont
  {A.}~\bibnamefont {Giudice}},\ and\ \bibinfo {author} {\bibfnamefont
  {A.}~\bibnamefont {Tosi}},\ }\bibfield  {title} {\bibinfo {title} {Low-noise
  ingaas/inp single-photon avalanche diodes for fiber-based and free-space
  applications},\ }\href {https://doi.org/10.1109/JSTQE.2021.3104962}
  {\bibfield  {journal} {\bibinfo  {journal} {IEEE Journal of Selected Topics
  in Quantum Electronics}\ }\textbf {\bibinfo {volume} {28}},\ \bibinfo {pages}
  {1} (\bibinfo {year} {2022})}\BibitemShut {NoStop}%
\bibitem [{\citenamefont {Ruggeri}\ \emph {et~al.}(2015)\citenamefont
  {Ruggeri}, \citenamefont {Ciccarella}, \citenamefont {Villa}, \citenamefont
  {Zappa},\ and\ \citenamefont {Tosi}}]{ruggeri2015integrated}%
  \BibitemOpen
  \bibfield  {author} {\bibinfo {author} {\bibfnamefont {A.}~\bibnamefont
  {Ruggeri}}, \bibinfo {author} {\bibfnamefont {P.}~\bibnamefont {Ciccarella}},
  \bibinfo {author} {\bibfnamefont {F.}~\bibnamefont {Villa}}, \bibinfo
  {author} {\bibfnamefont {F.}~\bibnamefont {Zappa}},\ and\ \bibinfo {author}
  {\bibfnamefont {A.}~\bibnamefont {Tosi}},\ }\bibfield  {title} {\bibinfo
  {title} {Integrated circuit for subnanosecond gating of ingaas/inp spad},\
  }\href {https://doi.org/10.1109/JQE.2015.2438436} {\bibfield  {journal}
  {\bibinfo  {journal} {IEEE Journal of Quantum Electronics}\ }\textbf
  {\bibinfo {volume} {51}},\ \bibinfo {pages} {1} (\bibinfo {year}
  {2015})}\BibitemShut {NoStop}%
\bibitem [{\citenamefont {IDQuantique}()}]{id221}%
  \BibitemOpen
  \bibfield  {author} {\bibinfo {author} {\bibnamefont {IDQuantique}},\
  }\href@noop {} {\bibinfo {title} {Id221 infrared single-photon detector}},\
  \bibinfo {howpublished}
  {\url{https://www.idquantique.com/resources/id221/}}\BibitemShut {NoStop}%
\bibitem [{\citenamefont {Boaron}\ \emph {et~al.}(2018)\citenamefont {Boaron},
  \citenamefont {Boso}, \citenamefont {Rusca}, \citenamefont {Vulliez},
  \citenamefont {Autebert}, \citenamefont {Caloz}, \citenamefont {Perrenoud},
  \citenamefont {Gras}, \citenamefont {Bussi\`eres}, \citenamefont {Li},
  \citenamefont {Nolan}, \citenamefont {Martin},\ and\ \citenamefont
  {Zbinden}}]{boaron2018secure}%
  \BibitemOpen
  \bibfield  {author} {\bibinfo {author} {\bibfnamefont {A.}~\bibnamefont
  {Boaron}}, \bibinfo {author} {\bibfnamefont {G.}~\bibnamefont {Boso}},
  \bibinfo {author} {\bibfnamefont {D.}~\bibnamefont {Rusca}}, \bibinfo
  {author} {\bibfnamefont {C.}~\bibnamefont {Vulliez}}, \bibinfo {author}
  {\bibfnamefont {C.}~\bibnamefont {Autebert}}, \bibinfo {author}
  {\bibfnamefont {M.}~\bibnamefont {Caloz}}, \bibinfo {author} {\bibfnamefont
  {M.}~\bibnamefont {Perrenoud}}, \bibinfo {author} {\bibfnamefont
  {G.}~\bibnamefont {Gras}}, \bibinfo {author} {\bibfnamefont {F.}~\bibnamefont
  {Bussi\`eres}}, \bibinfo {author} {\bibfnamefont {M.-J.}\ \bibnamefont {Li}},
  \bibinfo {author} {\bibfnamefont {D.}~\bibnamefont {Nolan}}, \bibinfo
  {author} {\bibfnamefont {A.}~\bibnamefont {Martin}},\ and\ \bibinfo {author}
  {\bibfnamefont {H.}~\bibnamefont {Zbinden}},\ }\bibfield  {title} {\bibinfo
  {title} {Secure quantum key distribution over 421 km of optical fiber},\
  }\href {https://doi.org/10.1103/PhysRevLett.121.190502} {\bibfield  {journal}
  {\bibinfo  {journal} {Phys. Rev. Lett.}\ }\textbf {\bibinfo {volume} {121}},\
  \bibinfo {pages} {190502} (\bibinfo {year} {2018})}\BibitemShut {NoStop}%
\bibitem [{\citenamefont {Rusca}\ \emph
  {et~al.}(2018{\natexlab{a}})\citenamefont {Rusca}, \citenamefont {Boaron},
  \citenamefont {Grünenfelder}, \citenamefont {Martin},\ and\ \citenamefont
  {Zbinden}}]{rusca2018finite}%
  \BibitemOpen
  \bibfield  {author} {\bibinfo {author} {\bibfnamefont {D.}~\bibnamefont
  {Rusca}}, \bibinfo {author} {\bibfnamefont {A.}~\bibnamefont {Boaron}},
  \bibinfo {author} {\bibfnamefont {F.}~\bibnamefont {Grünenfelder}}, \bibinfo
  {author} {\bibfnamefont {A.}~\bibnamefont {Martin}},\ and\ \bibinfo {author}
  {\bibfnamefont {H.}~\bibnamefont {Zbinden}},\ }\bibfield  {title} {\bibinfo
  {title} {Finite-key analysis for the 1-decoy state qkd protocol},\ }\href
  {https://doi.org/10.1063/1.5023340} {\bibfield  {journal} {\bibinfo
  {journal} {Applied Physics Letters}\ }\textbf {\bibinfo {volume} {112}},\
  \bibinfo {pages} {171104} (\bibinfo {year} {2018}{\natexlab{a}})},\ \Eprint
  {https://arxiv.org/abs/https://doi.org/10.1063/1.5023340}
  {https://doi.org/10.1063/1.5023340} \BibitemShut {NoStop}%
\bibitem [{\citenamefont {Rusca}\ \emph
  {et~al.}(2018{\natexlab{b}})\citenamefont {Rusca}, \citenamefont {Boaron},
  \citenamefont {Curty}, \citenamefont {Martin},\ and\ \citenamefont
  {Zbinden}}]{rusca2018security}%
  \BibitemOpen
  \bibfield  {author} {\bibinfo {author} {\bibfnamefont {D.}~\bibnamefont
  {Rusca}}, \bibinfo {author} {\bibfnamefont {A.}~\bibnamefont {Boaron}},
  \bibinfo {author} {\bibfnamefont {M.}~\bibnamefont {Curty}}, \bibinfo
  {author} {\bibfnamefont {A.}~\bibnamefont {Martin}},\ and\ \bibinfo {author}
  {\bibfnamefont {H.}~\bibnamefont {Zbinden}},\ }\bibfield  {title} {\bibinfo
  {title} {Security proof for a simplified bennett-brassard 1984
  quantum-key-distribution protocol},\ }\href
  {https://doi.org/10.1103/PhysRevA.98.052336} {\bibfield  {journal} {\bibinfo
  {journal} {Phys. Rev. A}\ }\textbf {\bibinfo {volume} {98}},\ \bibinfo
  {pages} {052336} (\bibinfo {year} {2018}{\natexlab{b}})}\BibitemShut
  {NoStop}%
\bibitem [{\citenamefont {Hayashi}\ and\ \citenamefont
  {Nakayama}(2014)}]{hayashi2014security}%
  \BibitemOpen
  \bibfield  {author} {\bibinfo {author} {\bibfnamefont {M.}~\bibnamefont
  {Hayashi}}\ and\ \bibinfo {author} {\bibfnamefont {R.}~\bibnamefont
  {Nakayama}},\ }\bibfield  {title} {\bibinfo {title} {Security analysis of the
  decoy method with the bennett–brassard 1984 protocol for finite key
  lengths},\ }\href {https://doi.org/10.1088/1367-2630/16/6/063009} {\bibfield
  {journal} {\bibinfo  {journal} {New Journal of Physics}\ }\textbf {\bibinfo
  {volume} {16}},\ \bibinfo {pages} {063009} (\bibinfo {year}
  {2014})}\BibitemShut {NoStop}%
\bibitem [{\citenamefont {Hwang}(2003)}]{hwang2003quantum}%
  \BibitemOpen
  \bibfield  {author} {\bibinfo {author} {\bibfnamefont {W.-Y.}\ \bibnamefont
  {Hwang}},\ }\bibfield  {title} {\bibinfo {title} {Quantum key distribution
  with high loss: Toward global secure communication},\ }\href
  {https://doi.org/10.1103/PhysRevLett.91.057901} {\bibfield  {journal}
  {\bibinfo  {journal} {Phys. Rev. Lett.}\ }\textbf {\bibinfo {volume} {91}},\
  \bibinfo {pages} {057901} (\bibinfo {year} {2003})}\BibitemShut {NoStop}%
\bibitem [{\citenamefont {Lo}\ \emph {et~al.}(2005)\citenamefont {Lo},
  \citenamefont {Ma},\ and\ \citenamefont {Chen}}]{lo2005decoy}%
  \BibitemOpen
  \bibfield  {author} {\bibinfo {author} {\bibfnamefont {H.-K.}\ \bibnamefont
  {Lo}}, \bibinfo {author} {\bibfnamefont {X.}~\bibnamefont {Ma}},\ and\
  \bibinfo {author} {\bibfnamefont {K.}~\bibnamefont {Chen}},\ }\bibfield
  {title} {\bibinfo {title} {Decoy state quantum key distribution},\ }\href
  {https://doi.org/10.1103/PhysRevLett.94.230504} {\bibfield  {journal}
  {\bibinfo  {journal} {Phys. Rev. Lett.}\ }\textbf {\bibinfo {volume} {94}},\
  \bibinfo {pages} {230504} (\bibinfo {year} {2005})}\BibitemShut {NoStop}%
\bibitem [{\citenamefont {Canale}(2014)}]{canale2014}%
  \BibitemOpen
  \bibfield  {author} {\bibinfo {author} {\bibfnamefont {M.}~\bibnamefont
  {Canale}},\ }\emph {\bibinfo {title} {Classical processing algorithms for
  Quantum Information Security}},\ \href@noop {} {Ph.D. thesis},\ \bibinfo
  {school} {Department of Information Engineering, University of Padova}
  (\bibinfo {year} {2014})\BibitemShut {NoStop}%
\bibitem [{\citenamefont {Boaron}\ \emph {et~al.}(2016)\citenamefont {Boaron},
  \citenamefont {Korzh}, \citenamefont {Houlmann}, \citenamefont {Boso},
  \citenamefont {Lim}, \citenamefont {Martin},\ and\ \citenamefont
  {Zbinden}}]{boaron2016detector}%
  \BibitemOpen
  \bibfield  {author} {\bibinfo {author} {\bibfnamefont {A.}~\bibnamefont
  {Boaron}}, \bibinfo {author} {\bibfnamefont {B.}~\bibnamefont {Korzh}},
  \bibinfo {author} {\bibfnamefont {R.}~\bibnamefont {Houlmann}}, \bibinfo
  {author} {\bibfnamefont {G.}~\bibnamefont {Boso}}, \bibinfo {author}
  {\bibfnamefont {C.~C.~W.}\ \bibnamefont {Lim}}, \bibinfo {author}
  {\bibfnamefont {A.}~\bibnamefont {Martin}},\ and\ \bibinfo {author}
  {\bibfnamefont {H.}~\bibnamefont {Zbinden}},\ }\bibfield  {title} {\bibinfo
  {title} {Detector-device-independent quantum key distribution: Security
  analysis and fast implementation},\ }\href
  {https://doi.org/10.1063/1.4960093} {\bibfield  {journal} {\bibinfo
  {journal} {Journal of Applied Physics}\ }\textbf {\bibinfo {volume} {120}},\
  \bibinfo {pages} {063101} (\bibinfo {year} {2016})}\BibitemShut {NoStop}%
\bibitem [{\citenamefont {Wengerowsky}\ \emph {et~al.}(2018)\citenamefont
  {Wengerowsky}, \citenamefont {Joshi}, \citenamefont {Steinlechner},
  \citenamefont {H{\"u}bel},\ and\ \citenamefont
  {Ursin}}]{wengerowsky2018entanglement}%
  \BibitemOpen
  \bibfield  {author} {\bibinfo {author} {\bibfnamefont {S.}~\bibnamefont
  {Wengerowsky}}, \bibinfo {author} {\bibfnamefont {S.~K.}\ \bibnamefont
  {Joshi}}, \bibinfo {author} {\bibfnamefont {F.}~\bibnamefont {Steinlechner}},
  \bibinfo {author} {\bibfnamefont {H.}~\bibnamefont {H{\"u}bel}},\ and\
  \bibinfo {author} {\bibfnamefont {R.}~\bibnamefont {Ursin}},\ }\bibfield
  {title} {\bibinfo {title} {An entanglement-based wavelength-multiplexed
  quantum communication network},\ }\href
  {https://doi.org/10.1038/s41586-018-0766-y} {\bibfield  {journal} {\bibinfo
  {journal} {Nature}\ }\textbf {\bibinfo {volume} {564}},\ \bibinfo {pages}
  {225} (\bibinfo {year} {2018})}\BibitemShut {NoStop}%
\bibitem [{\citenamefont {Ribezzo}\ \emph {et~al.}(2022)\citenamefont
  {Ribezzo}, \citenamefont {Zahidy}, \citenamefont {Vagniluca}, \citenamefont
  {Biagi}, \citenamefont {Francesconi}, \citenamefont {Occhipinti},
  \citenamefont {Oxenløwe}, \citenamefont {Lon\ifmmode~\check{c}\else
  \v{c}\fi{}arić}, \citenamefont {Cvitić}, \citenamefont
  {Stip\ifmmode~\check{c}\else \v{c}\fi{}ević}, \citenamefont
  {Pu\ifmmode~\check{s}\else \v{s}\fi{}avec}, \citenamefont {Kaltenbaek},
  \citenamefont {Ram\ifmmode~\check{s}\else \v{s}\fi{}ak}, \citenamefont
  {Cesa}, \citenamefont {Giorgetti}, \citenamefont {Scazza}, \citenamefont
  {Bassi}, \citenamefont {De~Natale}, \citenamefont {Cataliotti}, \citenamefont
  {Inguscio}, \citenamefont {Bacco},\ and\ \citenamefont
  {Zavatta}}]{ribezzo2022}%
  \BibitemOpen
  \bibfield  {author} {\bibinfo {author} {\bibfnamefont {D.}~\bibnamefont
  {Ribezzo}}, \bibinfo {author} {\bibfnamefont {M.}~\bibnamefont {Zahidy}},
  \bibinfo {author} {\bibfnamefont {I.}~\bibnamefont {Vagniluca}}, \bibinfo
  {author} {\bibfnamefont {N.}~\bibnamefont {Biagi}}, \bibinfo {author}
  {\bibfnamefont {S.}~\bibnamefont {Francesconi}}, \bibinfo {author}
  {\bibfnamefont {T.}~\bibnamefont {Occhipinti}}, \bibinfo {author}
  {\bibfnamefont {L.~K.}\ \bibnamefont {Oxenløwe}}, \bibinfo {author}
  {\bibfnamefont {M.}~\bibnamefont {Lon\ifmmode~\check{c}\else
  \v{c}\fi{}arić}}, \bibinfo {author} {\bibfnamefont {I.}~\bibnamefont
  {Cvitić}}, \bibinfo {author} {\bibfnamefont {M.}~\bibnamefont
  {Stip\ifmmode~\check{c}\else \v{c}\fi{}ević}}, \bibinfo {author}
  {\bibfnamefont {Z.}~\bibnamefont {Pu\ifmmode~\check{s}\else \v{s}\fi{}avec}},
  \bibinfo {author} {\bibfnamefont {R.}~\bibnamefont {Kaltenbaek}}, \bibinfo
  {author} {\bibfnamefont {A.}~\bibnamefont {Ram\ifmmode~\check{s}\else
  \v{s}\fi{}ak}}, \bibinfo {author} {\bibfnamefont {F.}~\bibnamefont {Cesa}},
  \bibinfo {author} {\bibfnamefont {G.}~\bibnamefont {Giorgetti}}, \bibinfo
  {author} {\bibfnamefont {F.}~\bibnamefont {Scazza}}, \bibinfo {author}
  {\bibfnamefont {A.}~\bibnamefont {Bassi}}, \bibinfo {author} {\bibfnamefont
  {P.}~\bibnamefont {De~Natale}}, \bibinfo {author} {\bibfnamefont {F.~S.}\
  \bibnamefont {Cataliotti}}, \bibinfo {author} {\bibfnamefont
  {M.}~\bibnamefont {Inguscio}}, \bibinfo {author} {\bibfnamefont
  {D.}~\bibnamefont {Bacco}},\ and\ \bibinfo {author} {\bibfnamefont
  {A.}~\bibnamefont {Zavatta}},\ }\bibfield  {title} {\bibinfo {title}
  {Deploying an inter-european quantum network},\ }\href
  {https://doi.org/https://doi.org/10.1002/qute.202200061} {\bibfield
  {journal} {\bibinfo  {journal} {Advanced Quantum Technologies}\ }\textbf
  {\bibinfo {volume} {n/a}},\ \bibinfo {pages} {2200061} (\bibinfo {year}
  {2022})}\BibitemShut {NoStop}%
\bibitem [{\citenamefont {Tosi}\ \emph {et~al.}(2012)\citenamefont {Tosi},
  \citenamefont {Acerbi}, \citenamefont {Anti},\ and\ \citenamefont
  {Zappa}}]{tosi2012ingaas}%
  \BibitemOpen
  \bibfield  {author} {\bibinfo {author} {\bibfnamefont {A.}~\bibnamefont
  {Tosi}}, \bibinfo {author} {\bibfnamefont {F.}~\bibnamefont {Acerbi}},
  \bibinfo {author} {\bibfnamefont {M.}~\bibnamefont {Anti}},\ and\ \bibinfo
  {author} {\bibfnamefont {F.}~\bibnamefont {Zappa}},\ }\bibfield  {title}
  {\bibinfo {title} {Ingaas/inp single-photon avalanche diode with reduced
  afterpulsing and sharp timing response with 30 ps tail},\ }\href@noop {}
  {\bibfield  {journal} {\bibinfo  {journal} {IEEE Journal of quantum
  electronics}\ }\textbf {\bibinfo {volume} {48}},\ \bibinfo {pages} {1227}
  (\bibinfo {year} {2012})}\BibitemShut {NoStop}%
\bibitem [{\citenamefont {Lucamarini}\ \emph {et~al.}(2018)\citenamefont
  {Lucamarini}, \citenamefont {Yuan}, \citenamefont {Dynes},\ and\
  \citenamefont {Shields}}]{lucamarini2018overcoming}%
  \BibitemOpen
  \bibfield  {author} {\bibinfo {author} {\bibfnamefont {M.}~\bibnamefont
  {Lucamarini}}, \bibinfo {author} {\bibfnamefont {Z.~L.}\ \bibnamefont
  {Yuan}}, \bibinfo {author} {\bibfnamefont {J.~F.}\ \bibnamefont {Dynes}},\
  and\ \bibinfo {author} {\bibfnamefont {A.~J.}\ \bibnamefont {Shields}},\
  }\bibfield  {title} {\bibinfo {title} {Overcoming the rate--distance limit of
  quantum key distribution without quantum repeaters},\ }\href
  {https://doi.org/10.1038/s41586-018-0066-6} {\bibfield  {journal} {\bibinfo
  {journal} {Nature}\ }\textbf {\bibinfo {volume} {557}},\ \bibinfo {pages}
  {400} (\bibinfo {year} {2018})}\BibitemShut {NoStop}%
\bibitem [{\citenamefont {Bechmann-Pasquinucci}\ and\ \citenamefont
  {Tittel}(2000)}]{bechmann2000quantum}%
  \BibitemOpen
  \bibfield  {author} {\bibinfo {author} {\bibfnamefont {H.}~\bibnamefont
  {Bechmann-Pasquinucci}}\ and\ \bibinfo {author} {\bibfnamefont
  {W.}~\bibnamefont {Tittel}},\ }\bibfield  {title} {\bibinfo {title} {Quantum
  cryptography using larger alphabets},\ }\href
  {https://doi.org/10.1103/PhysRevA.61.062308} {\bibfield  {journal} {\bibinfo
  {journal} {Phys. Rev. A}\ }\textbf {\bibinfo {volume} {61}},\ \bibinfo
  {pages} {062308} (\bibinfo {year} {2000})}\BibitemShut {NoStop}%
\bibitem [{\citenamefont {Vagniluca}\ \emph {et~al.}(2020)\citenamefont
  {Vagniluca}, \citenamefont {Da~Lio}, \citenamefont {Rusca}, \citenamefont
  {Cozzolino}, \citenamefont {Ding}, \citenamefont {Zbinden}, \citenamefont
  {Zavatta}, \citenamefont {Oxenl\o{}we},\ and\ \citenamefont
  {Bacco}}]{vagniluca2020efficient}%
  \BibitemOpen
  \bibfield  {author} {\bibinfo {author} {\bibfnamefont {I.}~\bibnamefont
  {Vagniluca}}, \bibinfo {author} {\bibfnamefont {B.}~\bibnamefont {Da~Lio}},
  \bibinfo {author} {\bibfnamefont {D.}~\bibnamefont {Rusca}}, \bibinfo
  {author} {\bibfnamefont {D.}~\bibnamefont {Cozzolino}}, \bibinfo {author}
  {\bibfnamefont {Y.}~\bibnamefont {Ding}}, \bibinfo {author} {\bibfnamefont
  {H.}~\bibnamefont {Zbinden}}, \bibinfo {author} {\bibfnamefont
  {A.}~\bibnamefont {Zavatta}}, \bibinfo {author} {\bibfnamefont {L.~K.}\
  \bibnamefont {Oxenl\o{}we}},\ and\ \bibinfo {author} {\bibfnamefont
  {D.}~\bibnamefont {Bacco}},\ }\bibfield  {title} {\bibinfo {title} {Efficient
  time-bin encoding for practical high-dimensional quantum key distribution},\
  }\href {https://doi.org/10.1103/PhysRevApplied.14.014051} {\bibfield
  {journal} {\bibinfo  {journal} {Phys. Rev. Appl.}\ }\textbf {\bibinfo
  {volume} {14}},\ \bibinfo {pages} {014051} (\bibinfo {year}
  {2020})}\BibitemShut {NoStop}%
\bibitem [{\citenamefont {Zahidy}\ \emph {et~al.}(2022)\citenamefont {Zahidy},
  \citenamefont {Ribezzo}, \citenamefont {Lazzari}, \citenamefont {Vagniluca},
  \citenamefont {Biagi}, \citenamefont {Occhipinti}, \citenamefont
  {Oxenl{\o}we}, \citenamefont {Galili}, \citenamefont {Hayashi}, \citenamefont
  {Antonelli}, \citenamefont {Mecozzi}, \citenamefont {Zavatta},\ and\
  \citenamefont {Bacco}}]{zahidy20224}%
  \BibitemOpen
  \bibfield  {author} {\bibinfo {author} {\bibfnamefont {M.}~\bibnamefont
  {Zahidy}}, \bibinfo {author} {\bibfnamefont {D.}~\bibnamefont {Ribezzo}},
  \bibinfo {author} {\bibfnamefont {C.~D.}\ \bibnamefont {Lazzari}}, \bibinfo
  {author} {\bibfnamefont {I.}~\bibnamefont {Vagniluca}}, \bibinfo {author}
  {\bibfnamefont {N.}~\bibnamefont {Biagi}}, \bibinfo {author} {\bibfnamefont
  {T.}~\bibnamefont {Occhipinti}}, \bibinfo {author} {\bibfnamefont {L.~K.}\
  \bibnamefont {Oxenl{\o}we}}, \bibinfo {author} {\bibfnamefont
  {M.}~\bibnamefont {Galili}}, \bibinfo {author} {\bibfnamefont
  {T.}~\bibnamefont {Hayashi}}, \bibinfo {author} {\bibfnamefont
  {C.}~\bibnamefont {Antonelli}}, \bibinfo {author} {\bibfnamefont
  {A.}~\bibnamefont {Mecozzi}}, \bibinfo {author} {\bibfnamefont
  {A.}~\bibnamefont {Zavatta}},\ and\ \bibinfo {author} {\bibfnamefont
  {D.}~\bibnamefont {Bacco}},\ }\bibfield  {title} {\bibinfo {title}
  {4-dimensional quantum key distribution protocol over 52-km deployed
  multicore fibre},\ }in\ \href
  {https://opg.optica.org/abstract.cfm?URI=ECEOC-2022-Th3C.6} {\emph {\bibinfo
  {booktitle} {European Conference on Optical Communication (ECOC) 2022}}}\
  (\bibinfo  {publisher} {Optica Publishing Group},\ \bibinfo {year} {2022})\
  p.\ \bibinfo {pages} {Th3C.6}\BibitemShut {NoStop}%
\bibitem [{\citenamefont {Da~Lio}\ \emph {et~al.}(2021)\citenamefont {Da~Lio},
  \citenamefont {Cozzolino}, \citenamefont {Biagi}, \citenamefont {Ding},
  \citenamefont {Rottwitt}, \citenamefont {Zavatta}, \citenamefont {Bacco},\
  and\ \citenamefont {Oxenl{\o}we}}]{da2021path}%
  \BibitemOpen
  \bibfield  {author} {\bibinfo {author} {\bibfnamefont {B.}~\bibnamefont
  {Da~Lio}}, \bibinfo {author} {\bibfnamefont {D.}~\bibnamefont {Cozzolino}},
  \bibinfo {author} {\bibfnamefont {N.}~\bibnamefont {Biagi}}, \bibinfo
  {author} {\bibfnamefont {Y.}~\bibnamefont {Ding}}, \bibinfo {author}
  {\bibfnamefont {K.}~\bibnamefont {Rottwitt}}, \bibinfo {author}
  {\bibfnamefont {A.}~\bibnamefont {Zavatta}}, \bibinfo {author} {\bibfnamefont
  {D.}~\bibnamefont {Bacco}},\ and\ \bibinfo {author} {\bibfnamefont {L.~K.}\
  \bibnamefont {Oxenl{\o}we}},\ }\bibfield  {title} {\bibinfo {title}
  {Path-encoded high-dimensional quantum communication over a 2-km multicore
  fiber},\ }\href {https://doi.org/10.1038/s41534-021-00398-y} {\bibfield
  {journal} {\bibinfo  {journal} {npj Quantum Information}\ }\textbf {\bibinfo
  {volume} {7}},\ \bibinfo {pages} {63} (\bibinfo {year} {2021})}\BibitemShut
  {NoStop}%
\bibitem [{\citenamefont {Bacco}\ \emph {et~al.}(2021)\citenamefont {Bacco},
  \citenamefont {Biagi}, \citenamefont {Vagniluca}, \citenamefont {Hayashi},
  \citenamefont {Mecozzi}, \citenamefont {Antonelli}, \citenamefont
  {Oxenl{\o}we},\ and\ \citenamefont {Zavatta}}]{bacco2021characterization}%
  \BibitemOpen
  \bibfield  {author} {\bibinfo {author} {\bibfnamefont {D.}~\bibnamefont
  {Bacco}}, \bibinfo {author} {\bibfnamefont {N.}~\bibnamefont {Biagi}},
  \bibinfo {author} {\bibfnamefont {I.}~\bibnamefont {Vagniluca}}, \bibinfo
  {author} {\bibfnamefont {T.}~\bibnamefont {Hayashi}}, \bibinfo {author}
  {\bibfnamefont {A.}~\bibnamefont {Mecozzi}}, \bibinfo {author} {\bibfnamefont
  {C.}~\bibnamefont {Antonelli}}, \bibinfo {author} {\bibfnamefont {L.~K.}\
  \bibnamefont {Oxenl{\o}we}},\ and\ \bibinfo {author} {\bibfnamefont
  {A.}~\bibnamefont {Zavatta}},\ }\bibfield  {title} {\bibinfo {title}
  {Characterization and stability measurement of deployed multicore fibers for
  quantum applications},\ }\href {https://doi.org/10.1364/PRJ.425890}
  {\bibfield  {journal} {\bibinfo  {journal} {Photon. Res.}\ }\textbf {\bibinfo
  {volume} {9}},\ \bibinfo {pages} {1992} (\bibinfo {year} {2021})}\BibitemShut
  {NoStop}%
\bibitem [{\citenamefont {Natarajan}\ \emph {et~al.}(2012)\citenamefont
  {Natarajan}, \citenamefont {Tanner},\ and\ \citenamefont
  {Hadfield}}]{natarajan2012superconducting}%
  \BibitemOpen
  \bibfield  {author} {\bibinfo {author} {\bibfnamefont {C.~M.}\ \bibnamefont
  {Natarajan}}, \bibinfo {author} {\bibfnamefont {M.~G.}\ \bibnamefont
  {Tanner}},\ and\ \bibinfo {author} {\bibfnamefont {R.~H.}\ \bibnamefont
  {Hadfield}},\ }\bibfield  {title} {\bibinfo {title} {Superconducting nanowire
  single-photon detectors: physics and applications},\ }\href
  {https://doi.org/10.1088/0953-2048/25/6/063001} {\bibfield  {journal}
  {\bibinfo  {journal} {Superconductor Science and Technology}\ }\textbf
  {\bibinfo {volume} {25}},\ \bibinfo {pages} {063001} (\bibinfo {year}
  {2012})}\BibitemShut {NoStop}%
\bibitem [{\citenamefont {Dauler}\ \emph {et~al.}(2014)\citenamefont {Dauler},
  \citenamefont {Grein}, \citenamefont {Kerman}, \citenamefont {Marsili},
  \citenamefont {Miki}, \citenamefont {Nam}, \citenamefont {Shaw},
  \citenamefont {Terai}, \citenamefont {Verma},\ and\ \citenamefont
  {Yamashita}}]{dauler2014review}%
  \BibitemOpen
  \bibfield  {author} {\bibinfo {author} {\bibfnamefont {E.~A.}\ \bibnamefont
  {Dauler}}, \bibinfo {author} {\bibfnamefont {M.~E.}\ \bibnamefont {Grein}},
  \bibinfo {author} {\bibfnamefont {A.~J.}\ \bibnamefont {Kerman}}, \bibinfo
  {author} {\bibfnamefont {F.}~\bibnamefont {Marsili}}, \bibinfo {author}
  {\bibfnamefont {S.}~\bibnamefont {Miki}}, \bibinfo {author} {\bibfnamefont
  {S.~W.}\ \bibnamefont {Nam}}, \bibinfo {author} {\bibfnamefont {M.~D.}\
  \bibnamefont {Shaw}}, \bibinfo {author} {\bibfnamefont {H.}~\bibnamefont
  {Terai}}, \bibinfo {author} {\bibfnamefont {V.~B.}\ \bibnamefont {Verma}},\
  and\ \bibinfo {author} {\bibfnamefont {T.}~\bibnamefont {Yamashita}},\
  }\bibfield  {title} {\bibinfo {title} {{Review of superconducting nanowire
  single-photon detector system design options and demonstrated performance}},\
  }\href {https://doi.org/10.1117/1.OE.53.8.081907} {\bibfield  {journal}
  {\bibinfo  {journal} {Optical Engineering}\ }\textbf {\bibinfo {volume}
  {53}},\ \bibinfo {pages} {081907} (\bibinfo {year} {2014})}\BibitemShut
  {NoStop}%
\bibitem [{\citenamefont {Divochiy}\ \emph {et~al.}(2008)\citenamefont
  {Divochiy}, \citenamefont {Marsili}, \citenamefont {Bitauld}, \citenamefont
  {Gaggero}, \citenamefont {Leoni}, \citenamefont {Mattioli}, \citenamefont
  {Korneev}, \citenamefont {Seleznev}, \citenamefont {Kaurova}, \citenamefont
  {Minaeva}, \citenamefont {Gol'tsman}, \citenamefont {Lagoudakis},
  \citenamefont {Benkhaoul}, \citenamefont {L{\'e}vy},\ and\ \citenamefont
  {Fiore}}]{divochiy2008superconducting}%
  \BibitemOpen
  \bibfield  {author} {\bibinfo {author} {\bibfnamefont {A.}~\bibnamefont
  {Divochiy}}, \bibinfo {author} {\bibfnamefont {F.}~\bibnamefont {Marsili}},
  \bibinfo {author} {\bibfnamefont {D.}~\bibnamefont {Bitauld}}, \bibinfo
  {author} {\bibfnamefont {A.}~\bibnamefont {Gaggero}}, \bibinfo {author}
  {\bibfnamefont {R.}~\bibnamefont {Leoni}}, \bibinfo {author} {\bibfnamefont
  {F.}~\bibnamefont {Mattioli}}, \bibinfo {author} {\bibfnamefont
  {A.}~\bibnamefont {Korneev}}, \bibinfo {author} {\bibfnamefont
  {V.}~\bibnamefont {Seleznev}}, \bibinfo {author} {\bibfnamefont
  {N.}~\bibnamefont {Kaurova}}, \bibinfo {author} {\bibfnamefont
  {O.}~\bibnamefont {Minaeva}}, \bibinfo {author} {\bibfnamefont
  {G.}~\bibnamefont {Gol'tsman}}, \bibinfo {author} {\bibfnamefont {K.~G.}\
  \bibnamefont {Lagoudakis}}, \bibinfo {author} {\bibfnamefont
  {M.}~\bibnamefont {Benkhaoul}}, \bibinfo {author} {\bibfnamefont
  {F.}~\bibnamefont {L{\'e}vy}},\ and\ \bibinfo {author} {\bibfnamefont
  {A.}~\bibnamefont {Fiore}},\ }\bibfield  {title} {\bibinfo {title}
  {Superconducting nanowire photon-number-resolving detector at
  telecommunication wavelengths},\ }\href
  {https://doi.org/10.1038/nphoton.2008.51} {\bibfield  {journal} {\bibinfo
  {journal} {Nature Photonics}\ }\textbf {\bibinfo {volume} {2}},\ \bibinfo
  {pages} {302} (\bibinfo {year} {2008})}\BibitemShut {NoStop}%
\bibitem [{\citenamefont {Provazn\'{i}k}\ \emph {et~al.}(2020)\citenamefont
  {Provazn\'{i}k}, \citenamefont {Lachman}, \citenamefont {Filip},\ and\
  \citenamefont {Marek}}]{provaznik2020benchmarking}%
  \BibitemOpen
  \bibfield  {author} {\bibinfo {author} {\bibfnamefont {J.}~\bibnamefont
  {Provazn\'{i}k}}, \bibinfo {author} {\bibfnamefont {L.}~\bibnamefont
  {Lachman}}, \bibinfo {author} {\bibfnamefont {R.}~\bibnamefont {Filip}},\
  and\ \bibinfo {author} {\bibfnamefont {P.}~\bibnamefont {Marek}},\ }\bibfield
   {title} {\bibinfo {title} {Benchmarking photon number resolving detectors},\
  }\href {https://doi.org/10.1364/OE.389619} {\bibfield  {journal} {\bibinfo
  {journal} {Opt. Express}\ }\textbf {\bibinfo {volume} {28}},\ \bibinfo
  {pages} {14839} (\bibinfo {year} {2020})}\BibitemShut {NoStop}%
\end{thebibliography}%

\end{document}